# Close Pairs as Probes of the Galaxy's Chemical Evolution

**Dany Vanbeveren and Erwin De Donder**

Astrophysical Institute, Vrije Universiteit Brussel, Pleinlaan 2, 1050 Brussels, Belgium

**Abstract** Understanding the galaxy in which we live is one of the great intellectual challenges facing modern science. With the advent of high quality observational data, the chemical evolution modeling of our galaxy has been the subject of numerous studies in the last years. However, all these studies have one missing element which is 'the evolution of close binaries'. Reason: their evolution is very complex and single stars only perhaps can do the job.

(Un)Fortunately at present we know that a significant fraction of the observed intermediate mass and massive stars are members of a binary or multiple system and that certain objects can only be formed through binary evolution. Therefore galactic studies that do not account for close binaries may be far from realistic. We implemented a detailed binary population in a galactic chemical evolutionary model. Notice that this is not something simple like replacing chemical yields. Here we discuss three topics: the effect of binaries on the evolution of $^{14}N$, the evolution of the type Ia supernova rate and the effects on the G-dwarf distribution, the link between the evolution of the r-process elements and double neutron star mergers (candidates of short gamma-ray burst objects).

## 1. Introduction

The ensemble of the physics of galaxy and star formation, the dynamics of stars and gas clouds in a galaxy and the details of stellar evolution from birth till death as function of chemical composition makes the chemo-dynamical evolutionary model (CEM) of a galaxy. During the last four decades many groups all over the world have constructed CEMs with varying degree of sophistication. All these CEMs have one property in common: although most of them account in some parameterized way for the effects of supernova (SN) of Type Ia, the effect of binaries on galactic stellar populations in general, on the chemical yields in particular is largely ignored without justification. Since 2000 our group in Brussels is studying in a systematic way the effects of intermediate mass and massive binaries on populations of stars and stellar phenomena and on CEM. Our CEM has been described in detail in an extended review (De Donder and Vanbeveren, 2004, further DV2004) and references therein.

Summarizing:

- To describe the formation and evolution of the Galaxy we use the model discussed by Chiappini et al. (1997) which is supported by recent results of hydrodynamical simulations (Sommer-Larsen et al., 2003).

- The star formation rate is a function of the local gravitational potential and thus also depends on the total surface mass density. We use the prescription of Talbot and Arnet (1975), updated by Chiosi (1980). The values of the different parameters that enter the formalism and that we use in our CEM were discussed in DV2004.

Stellar evolution:

- The chemical yields of single stars with an initial mass = [0.1 ; 120] $M_o$ are calculated from detailed stellar evolutionary computations where, in the case of massive stars, the most recent stellar wind mass loss rate prescriptions are implemented in the evolutionary code. Of course, the stellar evolutionary dataset depends on the initial metallicity Z and as far as massive stars are concerned not in the least on the effects of Z (read Fe) on the stellar wind mass loss rates. Our dataset corresponds to the case where the Luminous Blue Variable (LBV) mass loss rate is independent from Z whereas the red supergiant star, pre-LBV OB-type star and Wolf-Rayet star mass loss rates depend on Z as predicted by the radiatively driven wind theory (Kudritzki et al., 1989).

- Similarly as for the single stars, the chemical yields of close (=interacting) binary stars with an initial primary mass = ]1 ; 120] $M_o$ are based on an extended data set of detailed stellar evolutionary calculations of binaries where the evolution of both components is followed simultaneously. Notice that the yields are function of primary mass, binary mass ratio, binary period, the physics of the Roche lobe overflow (RLOF) and/or common envelope process, the physics of mass accretion, the effects of the supernova explosion on binary parameters. Obviously for the massive binaries the stellar wind effects are included similarly as in massive single stars.

- We explore the effects of different initial binary mass ratio (q) distributions, i.e. one that peaks at small q-values (Hogeveen, 1992, we use the letter H), at large q-values (Garmany et al, 1980, we use the letter G) or we assume a flat q-distribution.

- The initial binary period distribution is flat in the Log.

- A Salpeter type IMF for single stars and for binary primaries. We assume that the IMF-slope does not vary in time, which is more than sufficient in order to illustrate the basic conclusion of the present paper.

- To simulate the effects of SN Ia, we calculate the SNIa rate from first population synthesis principles using the single-degenerate (SD) model of Hashisu et al., 1996, 1999) and/or the double degenerate (DD) model of Iben and Tutukov (1984) and Webbink (1984) in combination with the merger timescale due to gravitational wave radiation. We like to remind the reader that consequently performing binary population synthesis which is linked to a physical SN Ia model is the ONLY method that gives scientific reasonable results

- The initial binary formation frequency enters the model as a free parameter. The binary frequency may not be confused with the number of stars in binaries. Suppose that we have 3 O-type stars, 1 single and the two others in 1 binary. Then the binary frequency is 50% but 2/3 of the O-type stars are binary members. Furthermore, the observed binary frequency is not necessarily the same as the binary frequency at birth (on the ZAMS). Stars which are born as binary components may become single stars during the evolution of the binary (due to the merger process or due to the SN explosion of the original companion). Any observed star sample contains the binaries, the single stars which are born as single and the single stars which became single but which were originally binary members. This means that the observed binary frequency is always a lower limit of the binary frequency at birth. The observationally confirmed O-type spectroscopic (thus close) binary frequency in young open clusters has been reviewed by Mermilliod (2001). The rates are as high as 80% (IC 1805 and NGC 6231), and can be as low as 14% (Trumpler 14). The massive close binary (MCB) frequency in Tr 16 is at least 50% (Levato et al., 1991) and in the association Sco OB2 it is at least 74% (Verschueren et al., 1996). The results listed above can be considered as strong evidence that the MCB frequency may vary among open clusters and associations. Mason et al. (1998) investigated the bright Galactic O-type stars and concluded that the observed spectroscopic binary frequency in clusters and associations is between 34% and 61%, among field stars between 20% and 50% and among the O-type runaways between 5% and 26% (a runaway is defined as a star with a peculiar space velocity larger than 30 km/s). However, as argued by the authors, these percentages may underestimate the true MCB frequency due to selection effects. The close systems that are missing in most of the observational samples are systems with periods larger than 40 days and mass ratios smaller than 0.4 which are obviously harder to detect. Mason et al. conclude that it cannot be excluded that we are still missing about half of all the binaries. Accounting for the observed percentages given above, we therefore conclude that the O-type MCB frequency may be very large. Vanbeveren et al. (1998a, b) investigated the bright B0-B3 stars (stars with a mass between 8 $M_o$ and 20 $M_o$) in the Galaxy and concluded that at least 32% are a primary of an interacting close binary. Again due to observational selection, using similar arguments as for the O-type binaries, the real B0-B3 binary frequency could be at least a factor 2 larger. Let us remark that the observations discussed above give us a hint about the MCB frequency in the solar neighborhood. However, whether or not this frequency is universal is a matter of faith. Indirect evidence about the binary frequency comes from population number synthesis (PNS) studies. Using all we know about single star and binary evolution, we can calculate the number of binaries of some type and compare it with the observed number. It is obvious to realize that predicted numbers depend on the adopted initial binary frequency; it can be concluded that in order to obtain general correspondence with observed numbers, the initial MCB frequency in the simulations must be very large (> 50%). Interestingly, De Donder and Vanbeveren (2002, 2003b) compared observed and theoretically predicted SN type Ia rates with PNS of intermediate mass single stars and binaries, and also in this case the adopted initial intermediate mass close binary frequency must be very large in order to obtain correspondence (see also section 2).

- The stable Roche lobe overflow phase in binaries is assumed to be quasi-conservative (all mass lost by the loser is accreted by the gainer). Notice that in most of the binaries with an initial primary mass larger than 40 $M_o$ the Roche lobe overflow phase is avoided due to the very large stellar wind mass loss during the Luminous Blue Variable phase of the primary (the LBV scenario as it was originally introduced by Vanbeveren, 1991).

- The common envelope and/or spiral-in phase is treated according to the prescription of Iben and Tutukov (1984) and Webbink (1984). In some cases these processes lead to the merger of the two binary components; in DV2004 we critically discussed possible consequences of the merger process of different types of binaries on CEM simulations.

- The effect of the SN explosion on the binary and binary orbital parameters is included by adopting a distribution of SN-asymmetries (expressed as the kick velocity that the compact SN remnant gets depending on the SN-asymmetry) and integrating over this distribution; the distribution of SN-asymmetries is based on the observed space velocity distribution of single pulsars which let us suspect that the distribution is $\chi^2$-like (or Maxwellian but with a tail extending to very large values) with an average kick-velocity = 450 km/s.

- Black hole formation happens when the initial mass of a single star > 25 $M_o$ and > 40 $M_o$ for an interacting binary component. Notice that the minimum-mass-difference is due to the fact that the evolution of the helium core in a single star differs from the one of a binary component who lost its hydrogen rich layers due to RLOF during hydrogen shell burning on the thermal timescale (for more details see DV2004). From first physical principles it is unclear whether or not BH formation in massive stars is accompanied by significant matter ejection. De Donder and Vanbeveren (2003) investigated the consequences on CEM simulations of the early Galaxy of the latter uncertainty. It was concluded that a CEM where it is assumed that all massive stars with an initial mass larger than 40 $M_o$ form BH's without matter ejection predicts a temporal $^{16}O$ evolution which is at variance with observations. These observations are much better reproduced when during the core helium burning a star loses mass by stellar wind (a Wolf-Rayet type wind) which depends on the stellar Fe-abundance as predicted by the radiatively driven wind theory whereas prior to or during BH formation on average 4 $M_o$ of $^{16}O$ should be ejected.

*Scope of the present paper*

The effect of binaries on the temporal evolution of the galactic chemical elements has been discussed in extenso in DV2004. Here we will focus on the effect of binaries on the SNIa rate, on the G-dwarf metallicity relation, on the temporal evolution of the double neutron star binary mergers and of the r-process elements and on the temporal evolution of $^{14}N$.

## 2. The supernova rates and the G-dwarf metallicity relation

Table 1 gives SNIa/SNIb ratios predicted with our CEM for different values of the initial binary frequency, different mass ratio distributions and for the two SNIa models (SD or DD). We prefer number ratios rather than absolute rates since they do not depend on the adopted model parameters of star formation and Galaxy formation. The observed ratio for spiral Galaxies ~1.6 (Cappellaro et al., 1999) and we conclude that there is a factor 2-3 discrepancy between the predicted and observed SNIa/SNIbc ratio independent from the adopted SNIa progenitor scenario, SD or DD. Since all SNIa's and a major part of the SNIbc's are formed by binaries, increasing the binary frequency doesn't help, at least if we assume that the binary frequency among massive and intermediate mass stars is the same. From present observations it cannot be concluded that SNIa's are exclusively produced by one type of binary systems and therefore SNIa's may form via the SD and the DD channel together. Assuming the latter significantly increases the SNIa rate and consequently brings the number ratio much closer to the observed value. If we look at the G-dwarf disk metallicity distribution which is shown in figure 1 we also find a better agreement with the observed distribution in case that both the SD and DD channel produce SNIa's. This result can be understood by the fact that the G-dwarf disk metallicity distribution is very sensitive by the iron evolution which is according to our simulations primarily determined by the SNIa rate (about 60-70 % of the total iron content in the Galaxy comes from SNIa's when they are formed by the SD and DD model and an initial binary frequency ≥ 50% is adopted).

**Table 1**: The predicted supernova number ratios SNIa/SNIbc. The used symbols have the following meaning: $f_b$ = initial total binary frequency, two mass ratio distributions labelled H and G (see text), SD = single degenerate model for SN Ia, DD = double degenerate model for SN Ia }

| $f_b$ | q-distr. | SNIa model | SNIa/SNIbc |
|---|---|---|---|
| 40% (70%) | H | SD | 0.5 (0.8) |
| 40% (70%) | H | DD | 0.4 (0.6) |
| 40% (70%) | G | SD | 0.2 (0.2) |
| 40% (70%) | G | DD | 0.7 (0.9) |

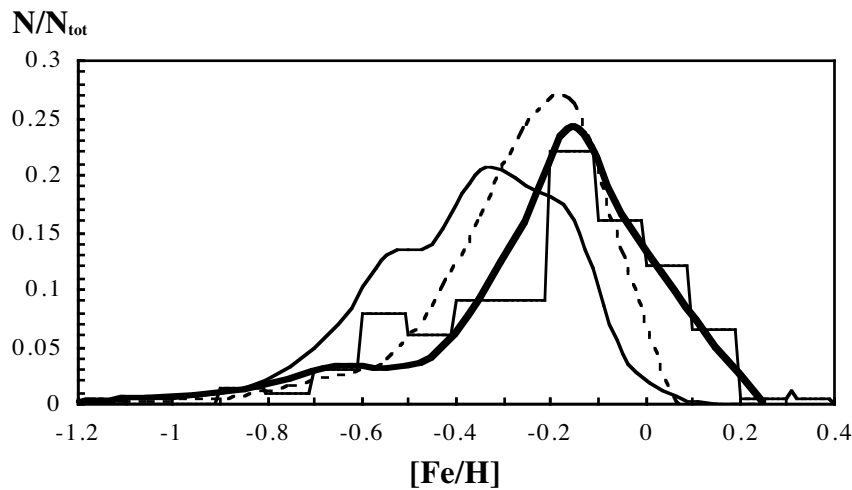

**Figure 1**: The predicted G-dwarf disk metallicity distribution when only the SD model is applied (full thin line), only the DD model (dashed thin line) and when both the DD and SD model produce SNIa's (thick line). The observed distribution (=histogram) is from Wyse and Gilmore (1995). }

### 3. The merger rate of double neutron star binaries and the evolution of r-process elements

Two sites have favorable physical conditions in order to be a major r-process source: the supernova explosion of a massive star (SN II or SN Ibc) and the binary neutron star merger (NSM). A discussion of the pro's and the contra's of both sites is given in Qian (2000) and Rosswog et al. (2000). At present, neither of the two can be promoted as the main enrichment source without reasonable doubt. The main argument against the NSM scenario is the following. Figure 2 shows the temporal evolution of the element europium of which the behavior can be considered as typical for the r-process elements. As can be noticed, europium is observed in stars which were born only a few 10 Myr after the formation of the Galaxy, when the Fe abundance [Fe/H] ~-3.1. Mathews et al. (1992) predicted the temporal evolution of the NSM-rate in the Galaxy and concluded that double neutron star merging starts too late, much later than [Fe/H] = -3.1. However, Mathews et al. used knowledge that was available at that time whereas they did not perform detailed binary population synthesis in combination with chemical evolution. We therefore repeated this study with our CEM (De Donder and Vanbeveren, 2003; see also DV2004). Our CEM calculates in detail the temporal evolution of the NS+NS binary population. The corresponding evolution of the NSM-rate is then determined using the theory of gravitational wave radiation. We predict a significant population of double NS binaries with small periods and very large eccentricities which merge within 1 Myr. This short merging timescale explains why at present no NS+NS binaries are observed with very large eccentricity, but they play a crucial role in the temporal evolution of the NSM-rate. This is demonstrated in Figure 2 where we compare the observed temporal evolution of europium with our predicted temporal evolution of the NSM rate. We conclude that it can not be excluded that NSMs are important sites of r-process element enrichment of the Galaxy.

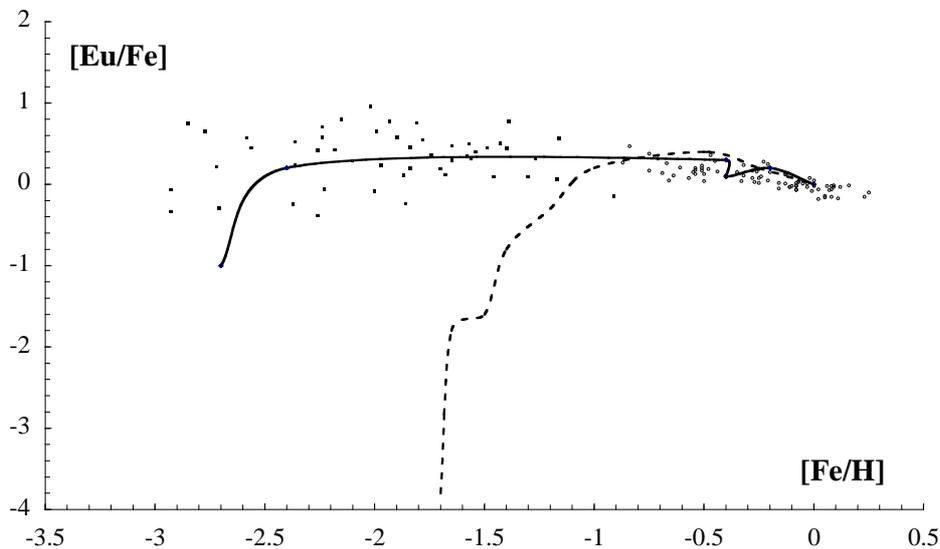

**Figure 2**: The temporal evolution of [Eu/Fe]. The squares are observations for halo stars from Burris et al. (2000), the circles are disk star observations from Woolf et al. (1995). On the same figure we also show the theoretically predicted NSM-rate as function of [Fe/H] (for the Solar neighborhood) (full line). The dashed line is calculated with the assumptions of Mathews et al. (1992). For both NSM-rate relations we used a normalization to assure that both lines explain the Solar abundance of Eu. We therefore had to assume that during an NSM-event on average 0.004 $M_o$ of Eu is ejected.

## 4. Very massive close binaries and the puzzling temporal evolution of $^{14}N$ in the Solar neighborhood

Reproducing the observed temporal evolution of $^{14}N$ during the early evolutionary phase of the Galaxy has been a problem since the beginning of chemical evolutionary modeling of Galaxies. This is illustrated in Figure 5 where we compare the observed $^{14}N$ evolution in the Solar neighborhood with our CEM prediction. In most of the CEMs it is assumed that the presently observed maximum stellar mass of stars in the Solar Neighborhood (=100-120 $M_o$) was the same in the past, i.e. independent from Z. The results of the Wilkinson Microwave Anisotropy Probe (WMAP) provide observational evidence that the low metallicity stellar initial mass function was top heavy with the possible presence of stars which were significantly more massive than 100-120 $M_o$ (the term *very massive star* is used). The optical depth along the line of sight to the last scattering surface of the Cosmic Microwave Background is interpreted in terms of the existence of a population of very massive stars in the early Universe (Kogut et al., 2003; Sokasian et al., 2003). Vanbeveren and De Donder (2006) investigated the possibility whether or not very massive stars can solve the $^{14}N$ problem. We first considered the following question: how massive can the maximum stellar mass be in order not to conflict with observations? Using the chemical yields of Heger and Woosley (2002) we adapted our CEM so that the effects of very massive stars during the early phases of Galaxy evolution can be studied. In Figure 3 we compare the observed temporal evolution of $^{12}$C with CEM predictions when it is assumed that the maximum stellar mass = 200 M$_o$ and when it is 260 M$_o$. With the latter value correspondence is very poor and we conclude that the maximum stellar mass during the early evolutionary phase of the Galaxy could have been 200 M$_o$ but not much larger. Figure 5 illustrates that very massive single star (with mass up to 200 M$_o$) can not solve the $^{14}$N discrepancy.

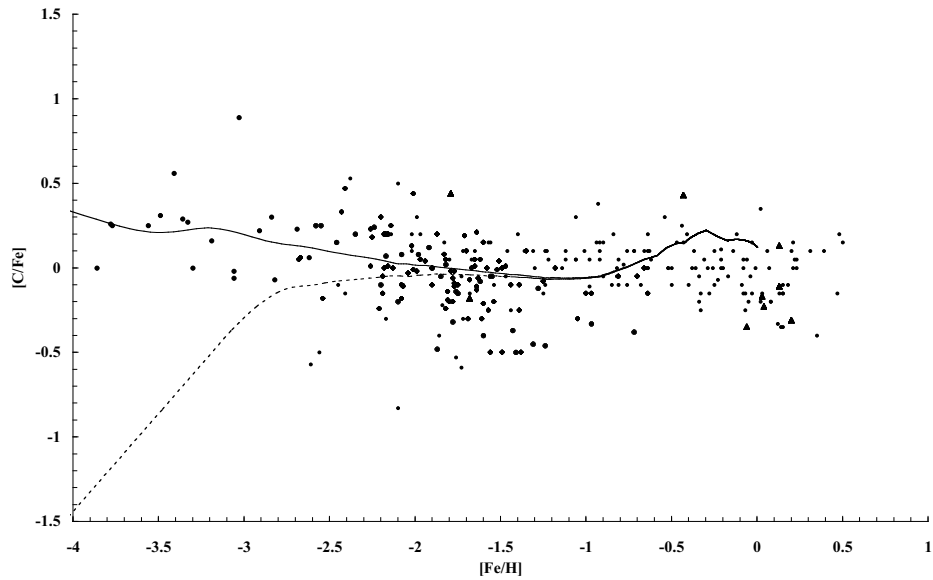

**Figure 3**: The observed (dots) versus the theoretically predicted temporal evolution of $^{12}$C. The thin line (resp. dashed line) corresponds to the case where very massive single stars with a mass $\leq$ 200 M$_o$ (resp $\leq$ 260 M$_o$) are included. The observations are from various sources which are discussed in De Donder and Vanbeveren (2004).

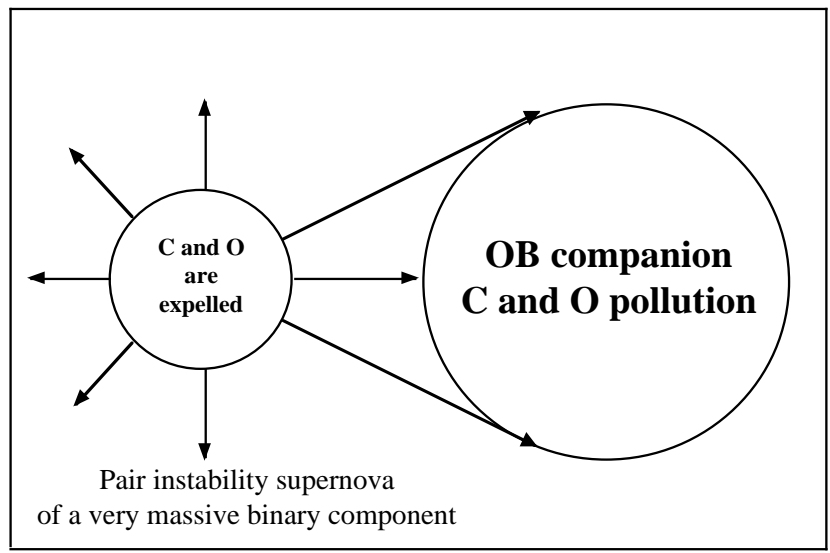

**Figure 4:** The very massive primary explodes as a pair instability SN and pollutes its companion.

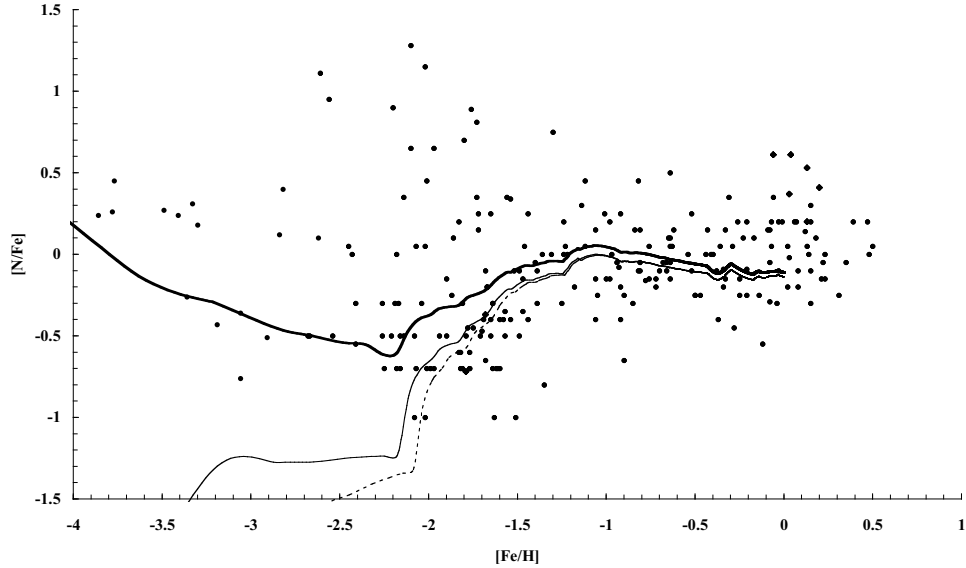

**Figure 5**: The observed (dots) versus the theoretically predicted temporal evolution of $^{14}N$. The dashed line is a CEM simulation without very massive stars. The thin line corresponds to the case where very massive single stars with a mass $\leq 200$ $M_o$ are included but no very massive binaries; the thick line corresponds to the case where we account for a 20% very massive binary frequency. The observations are from various sources which are listed and discussed in De Donder and Vanbeveren (2004).

Could very massive interacting binaries provide a way out? Let us consider a typical very massive binary consisting of a 90 $M_o$ He star (the post Roche lobe overflow remnant of a 180 $M_o$ very massive star) with a 140 $M_o$ companion. Low metallicity stars with a mass larger than 140 $M_o$ have nearly equal core hydrogen burning timescales. This means that when the He star explodes, our 140 $M_o$ is at the beginning of its core helium burning phase. Defining R as the radius of the companion, A as the binary separation, it is easy to understand that roughly

$$\frac{1}{4} 90 \frac{R^2}{A^2}$$

$M_o$ of the supernova matter (with a chemical composition which can be deduced directly from the yields published by Heger and Woosley, 2002) will be accreted by the companion (Figure 4 illustrates what happens). The evolutionary consequences are very similar to those during the Roche lobe overflow of a massive binary where nuclearly processed matter is transferred from the loser towards the gainer. The accreted matter has a molecular weight µ larger than the µ of the underlying layers which is an unstable configuration that initiates thermohaline mixing (Kippenhahn et al., 1980). Thermohaline mixing is a very rapid process and is capable to mix all layers outside the He-core (Braun, 1997). For the purpose of the present paper we treat the process as an instantaneous one and we mix the accreted mass in a homogeneous way with all the layers outside the He core. The foregoing instantaneous mixing process is strengthened by the fact that very massive core helium burning stars are supra-Eddington and most of the layers outside the He-core are convectively unstable. The core helium burning evolution of the companion after this mixing process is quite interesting. Depending on the period of the binary (thus on the value of A) the layers outside the He-core typically contain 0.3-0.5% (mass fraction) of $^{12}C$ (and about a factor of 10 more of $^{16}O$), thus also the layers where hydrogen burning takes place. The CN cycle in these layers rapidly transforms the available $^{12}C$ into $^{14}N$ which is ejected as primary nitrogen during the SN explosion. Our stellar evolutionary calculations for the 140 $M_o$ companion reveal that about 0.02-0.05 $M_o$ (depending on the binary period, thus on A) of $^{14}N$ are ejected, large enough to hope that it could solve the discrepancy between the observed and the theoretically predicted galactic temporal evolution of this element. To calculate whether or not very massive close binaries could explain the $^{14}N$ discrepancy, we adapted our standard CEM and we assumed that during the early evolutionary phase of the Galaxy 20% of all very massive stars are very massive close binaries which typically eject 0.05 $M_o$ of $^{14}N$, as discussed above. It can readily be understood that the $^{14}N$ yield predicted by the binary model discussed above scales lineary with the adopted binary frequency. A CEM where a 10% (resp. a 50%) very massive binary frequency is adopted would predict a $^{14}N$ yield that is a factor 2 lower (resp. a factor 2.5 larger) that the simulation with a 20% very massive binary frequency. Figure 5 also compares the observations with our prediction. We conclude:

> *within the uncertainties of the content of a population of very massive stars duringthe early evolution of our Galaxy, very massive close binaries can produce the observed earlytemporal evolution of $^{14}N$, if the binary component is polluted by the pair instability supernova ejecta of its companion and explodes.*

## 5. Summary

In the present paper we discussed the effects of binaries on galactic supernova rates, on the G-dwarf metallicity distribution, on and on the temporal evolution of $^{14}N$.

We conclude

- the observed SN Ia rate in spiral galaxies and the G-dwarf metallicity distribution of the Solar neighborhood is best reproduced by a CEM where SNIa are the result of the SD and DD together

- it can not be excluded that double neutron star mergers are important sites of r-process element enrichment of the Galaxy

- very massive close binaries with primary mass between 140 $M_o$ and 200 $M_o$ may be important production sites of $^{14}N$ during the early evolutionary phase of the Galaxy.